\documentclass[12pt]{article}
\usepackage{epsf}
\usepackage{graphicx}
\usepackage{psfig,color,epsfig}
\usepackage{eepic}
\usepackage{latexsym}
\usepackage{amsmath}
\newcommand{\be}{\begin{equation}}
\newcommand{\ee}{\end{equation}}
\newcommand{\ba}{\begin{array}}
\newcommand{\ea}{\end{array}}
\newcommand{\bqa}{\begin{eqnarray}}
\newcommand{\eqa}{\end{eqnarray}}
\begin{document}
%%%%%%%%%%%%%%%%%%%%%%%%%%%%%%%%%%%%%%%%%%%%%%%%%%%%%%%%%%%%%%
%%%%%%%%%%%%%%%%%%%%%%%%%%%%%%%%%%%%%%%%%%%%%%%%%%%%%%%%%%%%%
\begin{center}
{\Large\bf  Some Remarks on
 Exotic Resonances }\footnote{Talk  given at 10th International
  Symposium on Meson-Nucleon Physics and the Structure of the Nucleon (MENU 2004),
  Beijing, China, 29 Aug - 4 Sep 2004.
}
\\[1cm]
{\sc   Hanqing Zheng}
\\[5mm]
{\it  Department of Physics, Peking University, Beijing 100871,
P.~R.~China }
\\[0.5cm]
\begin{abstract}
Using large $N_c$ counting rule, it is argued that tetra-quark
resonances do not exist. Also it is pointed out that there exists
the violation of exchange degeneracy in the exotic $KN$ scattering
channel. It implies either the failure of resonance saturation
assumption or it suggests the existence of exotic baryon
resonances in such a channel.
\end{abstract}
\end{center}
Key words: tetraquark, pentaquark, large $N_c$, hadron--reggeon
duality

 \vspace{1cm}

There are revived theoretical interests  recently on the possible
existence of exotic resonances stimulated by various experimental
results on the pentaquark state $\Theta(1540)$\cite{pentaquark},
and the recently discovered narrow state
$D_{s,J}(2632)$\cite{SELEX}. The controversial $\Theta(1540)$
state, if established, will be the first observation of a
pentaquark state, while the  $D_{s,J}(2632)$ state are interpreted
as a tetraquark state by several groups\cite{tetraquark}.
Moreover, the width of the very broad meson resonance $f_0(600)$
(or $\sigma$) and $K_0^*(800)$ (or $\kappa$)\cite{PDG04} are
suggested to be non-vanishing in the large $N_c$
limit\cite{pelaez}, hence a non-conventional non $\bar qq$
state\cite{witten}, and probably a multi-quark state.

One effective way to study the problem of tetraquark state is to
use the effective Lagrangian approach. If a resonance decays into
pseudo-Goldstone bosons, it will unavoidably contribute to the low
energy constants of the effective chiral Lagrangian.\cite{GL84} To
illustrate the situation let us focus on the $\sigma$ resonance
appearing in I=0 $s$ wave $\pi\pi$ scatterings. We will use a
recently developed new parametrization form for partial wave $S$
matrix\cite{Kpi}, which has been proved to be very useful in
studying low lying broad resonances, for the low energy expansion
of resonances.\footnote{Or in the language of Lagrangian approch,
for integrating out resonances.}

In general the elastic partial wave $S$ matrix can be parametrized
as the following:\cite{Kpi}
 \be
 S^{phy.}=\prod_iS^{R_i}\cdot S^{cut}\ ,
\ee where $S^{R_i}$ denotes the `simplest' $S$ matrix with only
one pole or a pair of complex conjugate poles,\cite{Kpi} and
$S^{cut}$ denotes the cut contribution:
 \be
 S^{cut}=e^{2i\rho f(s)}\ ,\,\,\,  f(s)=\frac{s}{\pi}\int_{L+R}\frac{{\rm
 Im}_Lf(s')}{(s')(s'-s)}\  ,
\ee where $\mathrm{ Im}_Lf(s)=-\frac{1}{2\rho(s)}
\log|S^{phy.}(s)|$. Now, according to the conventional wisdom of
effective theory, the physical $S$ matrix, $S^{phy.}$ can be
faithfully represented by the $S$ matrix generated by chiral
perturbation theory ($\chi$PT) at low energies. Therefore, one can
match the low energy expansion of  $S^{R_i}$ with $\chi$PT
results. In this way one obtains the resonance contribution to the
low energy constants. The $N_c$ counting rules for the low energy
constants are standard\cite{GL84} and the resonance contribution
to the low energy constants has to obey such rules. One can
demonstrate that the $N_c$ counting rule for $f(s)$ is
$O(N_c^{-1})$,\cite{xiaotoappear} and therefore the matching can
be unambiguously made at leading order of 1/$N_c$ expansions. With
a little aid from micro-causality condition one reaches the
conclusion that no resonance of $M\sim O(1)$, $\Gamma\sim O(1)$
could exist. Especially the $\sigma$ resonance behaves like $M\sim
O(1)$, $\Gamma\sim O(1/N_c)$.

The above conclusion is only a special case of a more general
result. Similar conclusions should hold even when the tetra quark
state does not couple directly to pseudo-Goldstone bosons. The
picture is that, using Fierz transformations, it is easy to prove
that any quark anti-quark quadrilinear field can be decomposed as
products of quark anti-quark bilinear fields. That is to say a
tetra quark state will immediately fall apart into two meson pairs
if phase space is allowed. In other words, the width is $O(1)$ for
$N_c$ counting, but this is excluded because it will contribute to
the ordinary meson coupling vertices (or the low energy constants
in the Gasser--Leutwyler Lagrangian for pseudo-Goldstone bosons)
with wrongful $N_c$ order. If phase space is not allowed the
conclusion remains essentially the same, since the coupling
between a tetra-quark state and two normal mesons is still order
one.\footnote{There is one possibility that a quadrilinear bound
state escape the mismatch of the $N_c$ order, if it is accompanied
by a virtual bound state.\cite{xiaotoappear} Therefore, the
$f_0(980)$ can be consistent with a $\bar KK$ molecule
interpretation when and only when there exist a nearby virtual
state in the $\bar KK$ channel.} Therefore, {\it In the large
$N_c$ limit, quadrilinears make meson pairs and nothing
else.}\cite{coleman}

Next I will discuss one interesting topic concerning the old
concept of hadron--reggeon duality\cite{duality} in the exotic
channel of $KN$ scatterings.
 Assuming
isospin symmetry and charge conjugation invariance for the
reggeon--matter couplings, we write down the Regge amplitudes for
high energy $KN$ and ${\overline K}N$ elastic scatterings:
 \bqa
 T(K^-p\to K^-p)&=&P+\rho+\omega+f+A_2\ ,\nonumber\\
 T(K^+p\to K^+p)&=&P-\rho-\omega+f+A_2\ ,\nonumber\\
  T(K^+n\to K^+n)&=&P+\rho-\omega+f-A_2\ ,\nonumber\\
   T(K^-n\to K^-n)&=&P-\rho+\omega+f-A_2\ .
    \eqa
Also we have from isospin invariance the charge exchange
amplitude:
 \bqa T(K^+n\to K^0p)&=& T(K^+p\to K^+p)-T(K^+n\to
K^+n)\nonumber\\
&=&-2\rho+2A_2\ . \eqa
 For the reggeon exchange we have
 \be
  R=\beta_R(t)\frac{\eta
  -e^{-i\pi\alpha^R(t)}}{\sin(\pi\alpha^R(t))}\left(\frac{s}{s_0}\right)^{\alpha^R(t)}\
  ,
 \ee
 where $\eta$ is the signature factor, $\eta=-1$ for $P$, $f$ and
 $A_2$ while $\eta=+1$ for $\rho$ and $\omega$.
For the imaginary part of reggeon contribution, we have \be
\mathrm{Im}\,R=\beta_R\left(\frac{s}{s_0}\right)^{\alpha^R}\ . \ee
Optical theorem relates the $KN$ scattering total cross-section to
the imaginary part of the elastic scattering amplitude in the
forward direction, \be \sigma_{tot}(KN)=\frac{1}{2\cdot
\sqrt{\nu^2-(m_Nm_K)^2}} \mathrm{Im}\,T^{el}(s,t=0)\ ,\ee where
$\nu=p_1\cdot p_2$.

 Regge model affords an excellent
parameterization to the high energy scattering data. We repeat
such a fit for all $KN$ scattering data~\cite{PDG04} above
$s=10$GeV$^2$ (taking $s_0=1$GeV$^2$) and the results follow:
 \bqa
&& a^P=0.089\pm 0.001\ ,\beta^P=11.465\pm 0.070   \ ;\nonumber\\
&&a^R= 0.444\pm 0.009\ ,\nonumber\\
&&\beta^\rho=3.244\pm 0.177\ , \beta^{A_2}=1.906\pm
0.169\ ,\nonumber\\
&& \beta^\omega=10.86\pm  0.428\ ,\beta^f=24.06\pm 0.572\ .
 \eqa
 %The ffff fit quality is $\chi^2_{d.o.f.}=272/(209-7)$
 Strong exchange degeneracy requires $\beta^\rho=\beta^{A_2}$ but
 from above fit it is evident that there is a violation of SED in the exotic
 $KN$ scattering channels, though the
 effect is rather small. Such an effect  leaves an interesting possibility for
 the exotic resonances in the exotic channel according to the
 duality picture.

To make this point clearer we write down a $t=0$ finite energy sum
rule (FESR) for the $K^+n\to K^0p$ charge exchange scattering
amplitude,
 \bqa
&&\int^{L_2}_{-L_1}d\nu\,\nu^n\mathrm{Im}T(\nu,t=0)=\int^{L_2}_{-L_1}d\nu\,\nu^n
(\mathrm{Im}\,T_{K^+p\to K^+p}-\mathrm{Im}\,T_{K^+n\to
K^+n})\nonumber\\
&&=2\int^{L_2}_{-L_1}d\nu\,\sqrt{\nu^2-(m_Nm_K)^2}\nu^n(\sigma_t(K^+p)-\sigma_t(K^+n))\nonumber\\
&&=-2\frac{\beta_\rho-\beta_{A_2}}{\alpha(0)+n+1}L_2^{\alpha(0)+n+1}
-2\frac{\beta_\rho+\beta_{A_2}}{\alpha(0)+n+1}(-1)^nL_1^{\alpha(0)+n+1}\
.
 \eqa
The concept of hadron--reggeon duality suggests that reggeon
exchange represents an averaged effect of resonance contributions,
which means \bqa\label{fesr}
&&\int^{L_2}_{L_1}d\nu\,\nu^n\mathrm{Im}T_{K^+n\to K^0p}(\nu )
 =-2\frac{\beta_\rho-\beta_{A_2}}{\alpha(0)+n+1}(L_2^{\alpha(0)+n+1}-L_1^{\alpha(0)+n+1})
 \ ,\eqa
 where $L_1$ and $L_2$ are sufficiently large (much larger than 1 GeV), but arbitrary
 constants. If the left hand side of the above equation is saturated by the resonance contribution
 and if there is no resonance in the exotic channel at all then
 the $r.h.s.$ of the above equation must vanish. Fig.~1 illustrates the
 situation.
 \begin{figure}
\vspace{-2cm}
\centerline{\psfig{file=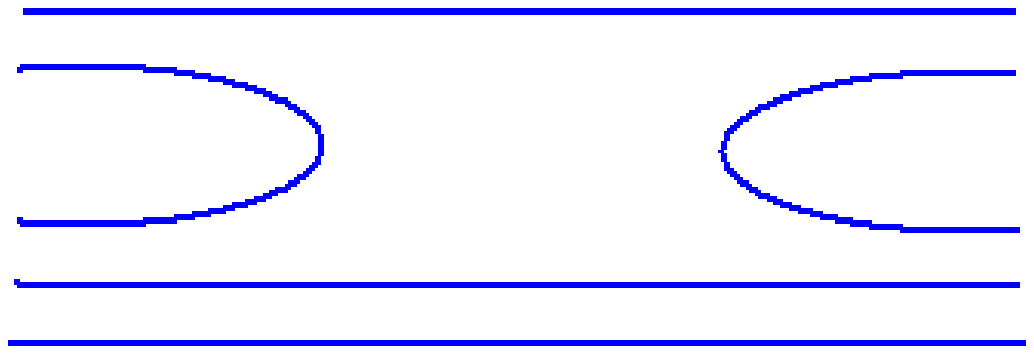,width=8cm}\psfig{file=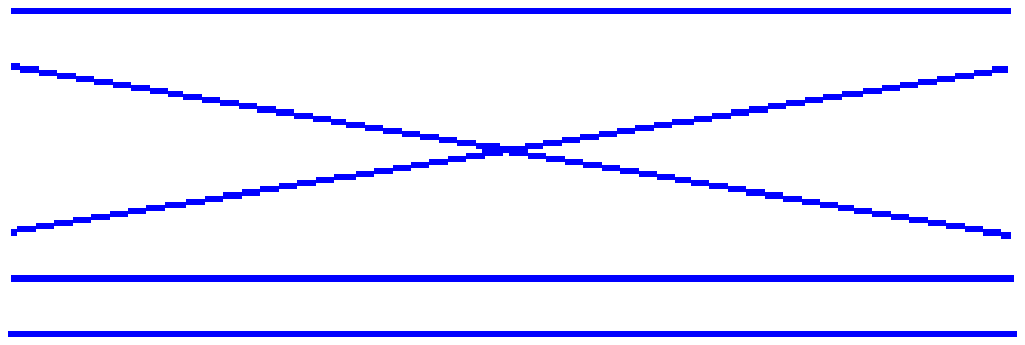,width=8cm}}
\vspace*{-4.5cm} \caption{Harari--Rosner duality diagram for
meson-baryon scatterings: left) non-exotic channel; right) exotic
channel. }
\end{figure}On the contrary, when there is a violation of exchange
 degeneracy (i.e.,$\beta_\rho\neq\beta_{A_2}$), it implies that there may
 exist exotic resonances. How to understand the fact that the exotic $KN$ scattering
 amplitude violates
strong exchange degeneracy? One possible explanation is that the
elegant duality picture is simply incorrect: in the exotic channel
reggeon exchange is dual to background rather than any resonance.
This problem remains to be open even if the $\Theta(1540)$ state
is proven to be illusive.

{\it Acknowledgement } I would like to thank Profs. Kuang-ta Chao
and  Shilin Zhu for helpful discussions.

\end{document}